# Ligand-induced Coupling versus Receptor Pre-association: Cellular automaton simulations of FGF-2 binding

**Manoj Gopalakrishnan*[1], Kimberly Forsten-Williams¶ and Uwe C. Täuber***

*\* Department of Physics, Virginia Polytechnic Institute and State University, Blacksburg, VA 24061-0435, USA.*

*¶ Department of Chemical Engineering, Virginia Polytechnic Institute and State University, Blacksburg, VA 24061-0211, USA.*

## Summary

The binding of basic fibroblast growth factor (FGF-2) to its cell surface receptor (CSR) and subsequent signal transduction is known to be enhanced by Heparan Sulfate Proteoglycans (HSPGs). HSPGs bind FGF-2 with low affinity and likely impact CSR-mediated signaling via stabilization of FGF-2-CSR complexes via association with both the ligand and the receptor. What is unknown is whether HSPG associates with CSR in the absence of FGF-2. In this paper, we determine conditions by which pre-association would impact CSR-FGF-2-HSPG triad formation assuming diffusion-limited surface reactions. Using mean-field rate equations, we show that (i) when [HSPG] is much higher than [CSR], the presence of pre-formed complexes does not affect the steady state of FGF-2 binding, and (ii) when the concentrations are comparable, the presence of pre-formed complexes substantially increases the steady state concentration of FGF-2 bound to CSR. These findings are supported by explicit cellular automaton simulations, which justify the mean-field treatment. We discuss the advantages of such a two-receptor system compared to a single receptor model, when the parameters are comparable. Further, we speculate that the observed high concentration of HSPG in intact cells ([HSPG] ~ 100[CSR]) provides a way to ensure that the binding levels of FGF-2 to its signaling receptor remains high, irrespective of the presence of pre-formed CSR-HSPG complexes on the cell surface, while allowing the cell to finely tune the response to FGF-2 via down-regulation of the signaling receptor.

---

[1] Corresponding author. E-mail: manoj@owl.phys.vt.edu, Tel: (540) 231-5344

# 1. Introduction

Growth factors play an important role in activating cellular processes and are critical for stimulating the proliferation of cells critical for wound healing and angiogenesis. Typically, growth factor activity is mediated by trans-membrane receptor proteins which transmit the chemical signals across the cell membrane. For many receptors, signal transduction requires both ligand binding and receptor dimerization, a process whereby two receptors either interact directly or are brought within close proximity to facilitate cross-reactivity or interaction with intracellular targets. Whether dimerization is exclusively ligand driven is still in question and may be dependent on the specific growth factor-receptor system.

While receptor dimerization appears to be a common paradigm, the receptor complex may consistent of different receptor proteins. For example, epidermal growth factor (EGF) and members of the EGF family can bind and activate both homo and hetero-dimers of the EGF receptor subfamily which includes EGFR, HER2, HER3, and HER4 (Hackel *et. al*, 1999). Basic fibroblast growth factor (bFGF or FGF-2) activates a trans-membrane cell surface receptor (CSR) via extra-cellular binding but also binds to the heparin sulfate glycosaminoglycan chains of cell surface proteoglycans. Although the interaction with heparan sulfate proteoglycans (HSPGs) is of lower affinity, interaction with these ubiquitous molecules has been shown to stabilize FGF-2-CSR binding and activation of CSR (Nugent and Edelman, 1992, Fannon &Nugent 1996). Although there is still some controversy over the exact stoichiometry of the signaling complex, the 2:2:2 model of 2 FGF-2, 2-HSPGs, and 2-CSR is most favored (Schlessinger *et. al.*, 2000, Plotnikov *et. al.*, 1999). The existence of preassembled CSR dimmers in the absence of FGF-2 is unlikely although whether CSR-HSPG complexes exist is not known. There is evidence that heparin and HSPGs can bind directly with the CSR (Powell, 2002; McKeehan, 1999; Kan, 1999) making such complexes possible. In this paper, we have focused on the formation of the FGF-2-CSR-HSPG triad given that this is likely the precursor to the signaling oligomer as is postulated for FGF-1 (Wu *et. al.*, 2003).

The purpose of this paper is to investigate the whether CSR-HSPG complexes, in the absence of FGF-2, would significantly impact formation of FGF-2-CSR-HSPG triads. Using a combination of mean-field rate equation analysis and direct numerical simulations, we demonstrate that when the HSPG concentration on the cell surface is much higher than that of CSR, as is normally found (reviewed in Tumova, 2000), the diffusion-limited surface reactions occur too fast as compared to the absorption of ligands from the bulk. In this case, the presence of pre-formed CSR-HSPG complexes makes little difference to the binding kinetics and the steady state. The situation when [HSPG] << [CSR] is also similar, except that the steady-state level of bound FGF-2 is then much lower compared to the previous case. However, pre-formed complexes become relevant when the ratio of these two concentrations is within a narrow range, typically 1-10. In this regime, the surface reactions are considerably slowed down due to the two-dimensional nature of the cell surface, and so the presence of pre-formed complexes significantly increases the steady-state level of FGF-2 bound to CSR. These results suggest a rationale for why the low-affinity receptor HSPG is present on the cell surface at such high concentrations as compared to the signaling receptor CSR. It might also be noted that still higher levels of HSPG could actually decrease the binding levels, since they could trap too much FGF-2, thus preventing their association with CSR.

Finally, we stress in this paper that dimerization of receptors is a general phenomena and, although the paper focuses on the FGF-2 system, the results are generalizable. We present results for the 1:2 growth factor:CSR system such as platelet-derived growth factor (PDGF) (reviewed in Claesson-Welch, 2000) and growth hormone (GH) (reviewed in Frank, 2002) and other factors.

## 2. Mathematical modeling

*1. Cellular automaton model of the FGF-2-CSR-HSPG system*

We have constructed a simple cellular automaton model of the FGF-2-CSR-HSPG system.

The model assumes that CSR and HSPG are distributed randomly on the two-dimensional cell surface with starting concentrations $R_0$ and $P_0$, respectively. The basic idea is to divide the cell surface into a grid array. A state index is then assigned to each site of this two-dimensional lattice, taking different values depending on whether the lattice site is vacant, or occupied by either a CSR or an HSPG molecule. A CSR molecule can be in three different states, free (R), bound to FGF-2 ($R'$), bound to HSPG ($R''$), or bound to both FGF-2 and HSPG ($R'''$). An HSPG molecule can be in two additional states, free (P) or bound to FGF-2 ($P'$). Thus, altogether, there are seven different states possible for a single lattice site.

The dynamics of the model take place at two stages, the absorption of FGF-2 from solution and the consequent binding and dissociation processes, and the various surface reaction processes; for example a FGF-2-bound CSR combines with a HSPG and forms a triad. Our first simplifying assumption is to treat the absorption of ligands from the bulk and the consequent binding to CSR and HSPG as an effective stochastic process, i.e., we do not study the diffusion of the ligand molecules in the solution explicitly. Rather, for a certain ligand concentration $\rho_L$, binding of FGF-2 onto R and P are modeled as first-order stochastic processes with time-independent rates as follows:

$R \rightarrow R'$ with probability $\alpha_R$ per unit time,
$P \rightarrow P'$ with probability $\alpha_P$ per unit time,
$R'' \rightarrow R'''$ with probability $\alpha_R$ per unit time.

The dissociation events are also described in a similar way:

$R' \to R$    with    probability $\beta_R$ per unit time,
$P' \to P$    with    probability $\beta_P$ per unit time,
$R''' \to R''$    with    probability $\gamma$    per unit time.

In the above, $\alpha_R = k_{on}^R \rho_L$, and $k_{on}^R$ represents the association rate of FGF-2 and CSR. The bulk ligand concentration $\rho_L$ is assumed to be independent of time. That is, we assert that the concentration of ligands is sufficiently high so that depletion of ligands in the solution is negligible. The local fluctuations in the ligand concentration in space and time are neglected henceforth. Similarly, $\alpha_P = k_{on}^P \rho_L$ where $k_{on}^P$ is the association rate of FGF-2 and HSPG. The association rates $k_{on}^R$ and $k_{on}^P$, as well as the dissociation rates $\beta_R$, $\beta_P$ and $\gamma$ used in our simulations were measured experimentally by Nugent and Edelman (1992) and are listed in Table 1.

The (second-order) rate constants for the surface reactions, however, have not been measured. There are three surface reactions to consider here. The first two are

$R' + P \xrightarrow{\lambda_D} R'''$,
$R + P' \xrightarrow{\lambda_D} R'''$,

which are the triad-forming reactions. The third surface reaction we consider is

$R + P \xrightarrow{\lambda_D} R''$,

whereby direct interaction between CSR and HSPG leads to the formation of a pre-formed hetero-dimer even in the absence of FGF-2. In the absence of experimental measurements, we assume that all these reactions are diffusion-limited, so that the rate constant are the same, which we denote by $\lambda_D$. The subscript D represents the diffusion coefficient of large proteins on the cell surface (which we assume identical for all species on the cell surface), and indicates that the reactions are diffusion-limited. In Appendix A, we show how this coefficient can be determined in the framework of Smoluchowski theory. The typical value of D for a membrane molecule is on the order of $10^{-9}$ cm$^2$/sec (Kucik *et. al.*, 1999).

The last reaction we consider is the dissociation of the pre-formed complex into CSR and HSPG,

$R'' \to R + P$.

We denote the dissociation constant for this reaction by g. There exist no experimental measurements of this rate to the best of our knowledge. Along with the diffusive association rate $\lambda_D$, g determines the fraction of CSR in the pre-formed CSR-HSPG state before ligand absorption, and features as a crucial control parameter in our computations. This calculation of this fraction, which we denote by $r^*(g)$, is detailed in Appendix B.

Our approach here is two-fold. In the following subsection, we further simplify our treatment by neglecting the spatial variation in the concentrations of all species. This leads to a set of first-order *mean-field* rate equations for the average concentrations of all quantities. From the rate equations, we derive the relations between all the steady-state concentrations, which are then solved, either analytically or numerically.

We construct a mean-field *steady state diagram* of the system where we identify the regimes where the majority of FGF-2-bound CSR is in either triad state or binary-complex state. The parameters that we use in the construction of this state diagram are (i) the ligand concentration (ii) the ratio of [HSPG] to [CSR], $n = P_0/R_0$ (iii) the absorption rate of FGF-2 and (iv) the surface reactivity $\lambda_D$. The last subsection presents results of direct numerical simulation of the cellular automaton model at a few key points in the steady state diagram. The results of the simulations confirm the results of the mean-field analysis, thus justifying the underlying assumptions. The simulations also bring to light some interesting features in the kinetics of the model. A brief discussion of these points will be the content of Sec. 4.

*2. Mean-field rate equations*

The complete set of mean-field rate equations for the time evolution of the concentrations [R], [R′], [R″], [R‴], [P] and [P′] are given below. (For simplicity, we shall henceforth omit the square brackets to denote the concentrations).

$$\frac{dR}{dt} = \beta_R R' + gR'' - \alpha_R R - \lambda_D R(P' + P), \tag{1}$$

$$\frac{dR'}{dt} = \alpha_R R - \beta_R R - \lambda_D R'P, \tag{2}$$

$$\frac{dR''}{dt} = \lambda_D RP - gR'' - \alpha_R R'' + \gamma R''', \tag{3}$$

$$\frac{dR'''}{dt} = \lambda_D (RP' + R'P) + \alpha_R R'' - \gamma R''', \tag{4}$$

$$\frac{dP}{dt} = \beta_p P' + gR'' - \lambda_D P(R' + R) - \alpha_p P, \tag{5}$$

$$\frac{dP'}{dt} = \alpha_p P - \beta_p P' - \lambda_D RP'. \tag{6}$$

We shall not attempt to solve these dynamical equations analytically or numerically. Rather, our approach here is to first determine the steady-state values of these quantities by putting the l.h.s. equal to zero and solving the resulting non-linear equations numerically, which we shall do in the next section. Our focus is to see how the steady-state level of the stable complex $R'''$ is affected by the initial concentration of the pre-formed hetero-dimer $R''$ for different ratios of $P_0$ to $R_0$. After identifying the relevant range of concentration, we run direct numerical simulations of the reaction-diffusion system to verify our predictions. For the sake of concreteness and ease of comparison with experimental data, we work with FGF-2 concentrations of 0.55 nM or its multiples in powers of 10 throughout this paper.

## 3. Model results

In this section, we outline the main results of our model.

*1. The steady state*

It is convenient to render the equations dimensionless by defining rescaled parameters via scaling all rates with $\alpha_R$. The rescaled variables are:

$$\tau = \alpha_R t, \alpha'_P = \frac{\alpha_P}{\alpha_R}, \beta'_R = \frac{\beta_R}{\alpha_R}, \beta'_P = \frac{\beta_P}{\alpha_R}, \gamma' = \frac{\gamma}{\alpha_R}, \lambda'_D = \frac{\lambda_D R_0}{\alpha_R} \text{ and } g' = \frac{g}{\alpha_R}.$$

We also define normalized concentrations: $r = \frac{R}{R_0}$, $r' = \frac{R'}{R_0}$, $r'' = \frac{R''}{R_0}$, $r''' = \frac{R'''}{R_0}$,

$p = \frac{P}{P_0}$ and $p' = \frac{P'}{P_0}$. Here $R_0$ and $P_0$ denote the initial concentrations of CSR and HSPG, respectively. Throughout this paper, we will work in the regime $P_0 \geq R_0$ ($n \geq 1$).

After applying these substitutions to eqns.(1-6), we arrive at the equivalent dimensionless set of equations

$$\frac{dr}{d\tau} = \beta'_R r' + g'r'' - r - \lambda'_D n(rp' + rp) ,\qquad(1a)$$

$$\frac{dr'}{d\tau} = r - \beta'_R r' - \lambda'_D nr'p ,\qquad(2a)$$

$$\frac{dr''}{d\tau} = \lambda'_D nrp - g'r'' - r'' + \gamma'r''' ,\qquad(3a)$$

$$\frac{dr'''}{d\tau} = \lambda'_D n(rp' + r'p) + r'' - \gamma'r''' ,\qquad(4a)$$

$$\frac{dp}{d\tau} = \beta'_P p' + g'r'' - \lambda'_D (r'p + rp) \quad \text{and} \qquad(5a)$$

$$\frac{dp'}{d\tau} = \alpha'_P p - \beta'_P p' - \lambda'_D r p' \ . \tag{6a}$$

To find the steady state of this set of rate equations, we put all time derivatives to zero. This yields a set of relations between the various concentrations (for simplicity of notation, we shall not use distinct symbols for the steady-state concentrations). From eqn. (6a), we find

$$p' = \frac{\alpha'_P p}{\beta'_P + \lambda'_D r'} \ , \tag{7}$$

while eqn.(2a) yields

$$r' = \frac{r}{\beta'_R + \lambda'_D np} \ , \tag{8}$$

and we obtain from eqn.(3a)

$$r'' = \frac{\gamma'}{1+g'} r''' + \frac{\lambda'_D}{1+g'} nrp \ . \tag{9}$$

Upon substituting eqns. (7), (8), and (9) into eqn. (4a), we arrive at

$$r''' = \frac{\lambda'_D npr(1+g')}{\gamma'g'} \left[ \frac{1}{1+g'} + \frac{\alpha'_P}{\beta'_P + \lambda'_D r} + \frac{1}{\beta'_R + \lambda'_D np} \right] \ . \tag{10}$$

We have thus expressed all the concentrations in terms of just two variables r and p. There are two additional constraints that relate these two, namely

$$r + r' + r'' + r''' = 1 \quad \text{and} \quad n(1 - p - p') = 1 - r - r' \ . \tag{11}$$

Eqn. (11) is just the dimensionless version of the normalization relations

$$R + R' + R'' + R''' = R_0 \quad \text{and} \quad P + P' + R'' + R''' = P_0 \ . \tag{12}$$

In general, eqns. (7-11) could be solved numerically to find the steady-state values of all the concentrations. However, before proceeding to do so, it is illuminating to study a simpler version of the model. Let us assume that the pre-formed complexes are absent in the model, i.e., there is no coupling between CSR and HSPG in the absence of FGF-2. This also means that when a triad releases FGF-2 back to the solution, both CSR and HSPG are liberated from the complex. The simplified set of reactions describing these processes is:

$$R \underset{\beta_R}{\overset{\alpha_R}{\rightleftarrows}} R' \ ,$$

$$P \underset{\beta_P}{\overset{\alpha_P}{\rightleftarrows}} P' \ ,$$

$$R' + P \xrightarrow{\lambda_D} R''' \xleftarrow{\lambda_D} R + P' \quad \text{and}$$

$$R''' \xrightarrow{\gamma} R + P \ .$$

The steady-state relations between the different concentrations in this simplified model become

$$p = \frac{\beta'_P + \lambda'_D r}{\beta'_P + \lambda'_D r + \alpha'_P} \ , \tag{14}$$

$$p' = \frac{\alpha'_P p}{\beta'_P + \lambda'_D r} \quad \text{and} \tag{15}$$

$$r''' = \frac{\lambda'_D n (rp' + r'p)}{\gamma'} \ . \tag{16}$$

After substituting eqns. (14) and (15) into eqn. (16), and using the normalization (11), we arrive at

$$\frac{r}{(\beta'_R + \lambda'_D np)} \left[ \beta'_R + \lambda'_D np + 1 + \frac{\lambda'_D n}{\gamma'} \left[ (\beta'_R + \lambda'_D np)(1-p) + p \right] \right] = 1 \ , \tag{17}$$

which, together with eqn.(13), implicitly determines r, and hence all the other concentrations. We note that in eqn. (17), the effective surface reaction rate is the product $\lambda'_D n$. This means that for such two-receptor systems, slow surface diffusion is compensated by increasing the imbalance between the receptor concentrations and vice-versa.

In general, the expression for the steady state value of r in this simplified model also can only be solved numerically, using e.g. the bisection method (Press *et al.,* 1990*)*. However, explicit expressions for all concentrations may be obtained when the ligand concentration is sufficiently high such that the rates $\beta'_P, \beta'_R, \gamma', \lambda'_D \to 0$ (recall that these dimensionless rates are inversely proportional to $\alpha_R$, and therefore to $\rho_L$). Only $\alpha'_P$ remains constant in this limit (because both $\alpha_R$ and $\alpha_P$ are proportional to $\rho_L$), whence we obtain the following simplified expressions for r, r' and r''':

$$r \approx \frac{\beta'_R}{\alpha'_P + \frac{\lambda'_D n}{\gamma'}(\beta'_R + \frac{\beta'_P}{\alpha'_P})} \to 0 \quad \text{for large } \rho_L, \tag{18}$$

$$r' \approx \frac{\alpha'_P}{\alpha'_P + \frac{\lambda'_D n}{\gamma'}(\beta'_R + \frac{\beta'_P}{\alpha'_P})} \to 1 \quad \text{for large } \rho_L \text{ and} \tag{19}$$

$$r''' \approx \frac{\frac{\lambda'_D n}{\gamma'}(\beta'_R + \frac{\beta'_P}{\alpha'_P})}{\alpha'_P + \frac{\lambda'_D n}{\gamma'}(\beta'_R + \frac{\beta'_P}{\alpha'_P})} \to 0 \quad \text{for large } \rho_L. \tag{20}$$

We see that for fixed n and large $\rho_L$, the concentration of the simple CSR-FGF-2 complexes increases at the expense of the triad CSR-FGF-2-HSPG. This is reasonable because at large ligand concentrations, most of CSR and HSPG will be bound to FGF-2 in a short time, so that there is no appreciable triad formation (in the absence of pre-formed CSR-HSPG complexes). However, if this ratio n also increases in proportion to $\rho_L$, the triad concentration remains constant. These observations are summarized in a steady state diagram for the model in Fig.1. Upon increasing $\rho_L$ for fixed n, the concentration of the binary complex FGF-2-CSR increases at the expense of the triad, eventually dominating it at sufficiently high ligand concentration. We also see that as n increases, the imbalance in the reactants makes the dynamics more favorable to triad formation, so the threshold ligand concentration required to enter the binary complex dominated regime also increases proportional to n.

It is instructive at this point to see what a similar steady state diagram would look like for a ligand-receptor system with only one type of receptor (which, we again designate as R). Dimerization of a single receptor type (in the absence of stabilizing low affinity receptors) for growth factor activity is the paradigm for many growth factors. The mean-field calculation for this simple system is very similar to the two-receptor case we have studied so far, and has the advantage that the complete solution can be found explicitly (again, in the absence of pre-formed R-R dimers). The relevant calculations for this model is presented in Appendix C.

The steady state diagram depicted in Fig. 2 (which follows from eqn. C.9) shows that the R-L-R triad complex is the dominant form for bound R at high values of the diffusion coefficient and low values of the ligand absorption rate. For comparison, we have also shown the corresponding diagram for the original CSR-FGF-2-HSPG system in the same figure. In general, when everything else is the same, the threshold value of surface reaction rate for dominance of the triad is seen to be lower for the two-receptor system, and is still lowered as the imbalance ratio n is increased. It is thus seen that when the surface reactivity is small, the two-receptor system can effectively compensate for it by increasing the ratio of [HSPG] to [CSR]. It is also obvious that the threshold surface reactivity is inversely proportional to the initial concentration of CSR in both cases (which simply follows from the dimensionless scaling form of $\lambda_D$).

It is also interesting to look at the effect of changing the dissociation rates in the mean-field steady state diagram. This is most clearly seen for the simpler single-receptor model studied in Appendix C. For this case, it is seen from eqn. C.9 that the slope of the boundary between triad-dominated and binary-complex-dominated regimes is inversely proportional to the dissociation constant $\beta$. This means that, for higher $\beta$, the boundary line between the two regimes becomes flatter, since the binary complex is less stable now, for given values of $\alpha$ and $\lambda_D$. Thus the binary-complex-dominated regime is pushed down in the steady state diagram. Increasing the triad dissociation constant $\gamma$ would reduce the triad concentration, and hence has the opposite effect. At higher $\gamma$, both the y-intercept and the slope of the boundary becomes larger in magnitude, thus pushing the triad-dominated regime upwards in the steady state diagram. The mathematical expression of this behavior is easily seen from eqn. C.9: the y-intercept is proportional to $\gamma$, whereas the slope is proportional to $\gamma/(\beta+\gamma)$. The effects of increasing the dissociation rates on the two-receptor system are similar (Fig. 3).

The insights obtained from studying the simplified (no pre-formed complex) version of the model leads us to conjecture on the role of pre-formed CSR-HSPG complexes, which is the main purpose of this paper. In general, for both the single-receptor model and the CSR-FGF-2-HSPG system, the presence of pre-formed CSR-CSR (for single receptor model) or CSR-HSPG (for two-receptor model) complexes increases the triad concentration, and hence, the effect of such pre-coupling is significant only at the parameter values corresponding to the binary-complex-dominated regime of the steady state diagrams. In this regime, the presence of pre-formed complexes will boost the triad concentration, which would otherwise be small. This is exactly what we find from numerical solutions of the complete mean-field steady-state equations, eqn. 7-11. Fig.4 shows how the fraction of triads in the steady state varies with the fraction of CSR-HSPG complexes in initial state, for two different ligand concentrations, at n=1. The effect of pre-formed complexes is appreciable only at ligand concentrations substantially higher than typical experimental concentrations (which are of the order of 1-10 nM). In Fig.5, we show similar results, but upon increasing [HSPG], keeping the ligand concentration and [CSR] fixed. In this case, we find that the effect of pre-formed complexes is noticeable only when n is sufficiently low.

To conclude this section, the mean-field analysis has shown that pre-formed complexes are irrelevant to the system at low ligand concentrations and/or high values of [HSPG] to [CSR] ratio. These findings are supported through numerical simulations of the cellular automaton model, to be discussed in the subsequent section.

*2. Numerical simulations of the cellular automaton model*

We now proceed to our direct numerical Monte Carlo (MC) simulation of the full cellular automaton (CA) model. We divide the cell surface into a square lattice of $L \times L$ sites. The lattice spacing $\Delta$ is fixed as ¼ of the mean separation between two HSPG molecules. If $N_P$ is the number of HSPG molecules per cell, this gives

$$\Delta = \frac{1}{4} \frac{\xi}{\sqrt{N_P}} \quad , \tag{21}$$

where $\xi \approx 5\mu m$ is the typical linear dimension of the cell. We determine the initial concentration $R_0$ using an estimate of the number of receptor proteins per cell, which are approximately 15,000 for the FGF-2 system we are modeling (Nugent&Edelman, 1992).

The dynamics of the model is defined as follows. There are essentially two time scales in the problem. The characteristic time of absorption of ligands from the bulk is of the order of minutes for typical ligand concentrations used in binding experiments. On the other hand, the characteristic time scale of surface diffusion is the typical time it takes for diffusion of one molecule to another, which is of the order of one lattice spacing as introduced above. This time scale is thus given by

$$\tau^* = \frac{\Delta^2}{4D} \quad , \tag{22}$$

which represents one Monte Carlo time step for surface diffusion, and is the microscopic time scale in the problem.

Next, we need to choose the time scale for adsorption of ligand molecules from the bulk solution. For computational efficiency, we would like to define a time scale sufficiently large so as to ensure a reasonable number of absorption events at every MC step. At the same time, this period should be sufficiently small so as to not exceed the microscopic time scale $\tau^*$ too much. These considerations lead us to choose the time scale $\tau$ such that

$$\tau\omega = 0.1, \quad \omega = \max(\alpha_R, \alpha_P, \beta_R, \beta_P, \gamma, g) \quad . \tag{23}$$

In the simulation of the model, we implement the following procedure. We measure time in units of $\tau$. Thus, at each time step, we update the state of every site in the lattice, selected at random by attaching ligand molecules to CSR and HSPG occupied sites with certain probabilities. For each such absorption time step, we now perform $N = \frac{\tau}{\tau^*}$ time steps of surface diffusion, wherein one of the following events takes place. We choose one lattice site randomly, and select one of the neighboring sites also randomly. The rules used for updating the state of each site are as follows:

(i) if the selected site is occupied, and the chosen neighbor is vacant, then the site exchanges its state with the neighbor (diffusion);

(ii) if the site is occupied by CSR, and the neighbor is occupied by FGF-2-HSPG, then the CSR is replaced by CSR-FGF-2-HSPG, and the neighbor becomes vacant (reaction);

(iii) if this site is occupied by HSPG and the neighbor is occupied by FGF-2-CSR, then the state of the neighbor is updated to CSR-FGF-2-HSPG and the site is now vacant (reaction).

Following N diffusion time steps, we perform the dissociation events now, which are similar to the absorption events. This completes one MC time step of the simulation. All our simulations were done on a square lattice of size L=256 with periodic (toroidal) boundary conditions. The codes for the simulations were written in the FORTRAN 77 programming language and the jobs were run on a Dell PC in a Linux environment. For the generation of the random numbers, we used the *ran2* subroutine from *Numerical Recipes in Fortran* (Press *et. al.*, 1990). Although the lattice size used is much smaller than the size required for covering the typical cell area (judging from the lattice size chosen in eqn. 21), we chose it for computational efficiency. Moreover, because of the local imbalance in the concentrations of reactants in our model, typical reaction time scales in our problem are much larger than diffusion time scales, rendering finite size effects largely absent in our simulations.

We now proceed to discuss our numerical results. The simulations were executed in the regime where the concentrations of CSR and HSPG are nearly equal. When [HSPG] is very large compared to [CSR], the lattice spacing in the cellular automaton model is of the order of the typical diameter of these proteins. In such a situation, our simple model where each lattice site (occupied by CSR or HSPG) absorbs or releases FGF-2, independent of the state of its neighbors, is no longer valid. Most of the simulations we discuss below were done at ligand concentrations far higher than what is usually used in experiments ($\mu M$ instead of nM), since appreciable effects of pre-formed complexes were found only at these high ligand concentrations.

The numerical simulation data exhibits behavior qualitatively similar to that predicted by the mean-field steady state diagram (Figs.1 and 2). However, the typical time scales in the simulation to reach the steady state levels was found to be far smaller than what was found in experiments. We have included a discussion of this issue in the last section of the paper. In Fig. 6, we see that in the absence of direct CSR-HSPG interaction, the fraction of CSR in the triad state increases with increasing ligand concentration. However, even for ligand concentrations up to 5.5 $\mu M$ (four orders of magnitude higher than typical experimental values), we do not enter the regime where the binary complex dominates over the triad. One explanation for this behavior could be that, in the real system, the local imbalance in the concentrations of FGF-2-CSR and HSPG drives the triad concentration to a steady-state level higher than that predicted by the mean-field rate equations. Of course, it is also important to note that oligomerization of these triads to facilitate CSR-CSR cross-talk would also impact the overall dynamics of the experimental system.

In Fig. 7 and 8, we present numerical results where we have included direct coupling between CSR and HSPG. In other words, we let the CSR-HSPG complexes form through diffusion-limited reactions between CSR and HSPG prior to ligand absorption. The decay constant of the triad g is now tuned so as to convert any desired fraction of CSR into the CSR-HSPG state. Fig. 7 shows how the steady state level of triad concentration changes with the fraction of pre-formed complexes, when the concentrations of CSR and HSPG are equal (i.e., we have chosen n=1, where the change is maximum). The rate of change in the triad steady-state concentration as we increase the fraction of pre-formed complexes is seen to be larger at higher ligand concentration. For example, at ligand concentration $5.5\mu M$, if 60% of CSR is in the pre-formed complex state, the fraction of the triad complexes increases by a factor of nearly 1.5. At lower concentrations, more in keeping with experimental procedures, the effect is negligible. In Fig. 8, we have shown similar data, but by varying the [HSPG] to [CSR] ratio n, keeping the ligand concentration fixed. We observe that for a fixed ligand concentration, the fraction of triads increases with n. However, the change is noticeable only at very high ligand concentrations. We have further observed that increasing $R_0$ increases the effective surface reaction rate (note the scaling form for $\lambda_D$ discussed prior to eqn. 1a), and, in general, makes the situation more favorable to triad formation.

To conclude this section, we found that the results of Monte Carlo simulations are in qualitative agreement with the mean-field calculations. In general, the sensitivity of the triad concentration to the presence of pre-formed CSR-HSPG complexes increases with the association rate (or ligand concentration) and decreases with n, the ratio of [HSPG] to [CSR]. At low ligand concentration, more characteristic of what is used experimentally, the surface reactions occur much faster than absorption of ligands from bulk (using typical estimates for the diffusion coefficient of proteins in membranes, Kucik et al, 1991), so that the presence of pre-formed complexes is not important for the dynamics of the system. At high values of n, the local imbalance in the reactants accelerates the surface reactions too much, and thus, once again, pre-formed complexes become irrelevant.

## 4. Discussion and conclusions

In this paper, we have studied the FGF-2-CSR-HSPG ligand-receptor system both through analyzing the mean-field rate equations and by means of cellular automaton simulations. Experimentally, the presence of HSPG molecules has been shown to slow down the release of FGF-2 from its signaling receptor CSR by almost a factor of 10, while not significantly changing the association rate (Nugent and Edelman, 1992). HSPG, as well as heparin, also bind FGF-2, but with lower affinity than CSR, and the general consensus is that that HSPG stabilizes the CSR-FGF-2 complex. Like many other growth factors, receptor dimerization is thought to be necessary for signal transduction and stabilization by HSPG appears necessary for activity under physiological conditions (Fannon & Nugent, 1996).

Our basic purpose in this paper is to examine two possible mechanisms for the formation of the CSR-FGF-2-HSPG triad complexes - (a) the diffusion-limited reactions between CSR-FGF-2 and HSPG and between FGF-2-HSPG and CSR, and (b) existence of the pre-formed hetero-dimers CSR-HSPG on the cell surface. We sought to determine the conditions under which one of these mechanisms would have an advantage over the other.

At the purely mean-field level, where all the spatial organization of the reacting species is ignored, we predicted that for any value of the ratio n of initial concentrations of HSPG and CSR, there is a threshold value of ligand concentration, above which the CSR-FGF-2 complex would dominate over the triad in the absence of pre-formed CSR-HSPG complexes. This threshold value was found to increase linearly with n, which is the ratio of initial concentrations of HSPG and CSR. It may be noted that the experimentally measured value of n is cell dependent with a range between ~10 and 300 (Moscatelli, 1987) and the range mediated by changes in the level of CSR. The threshold ligand concentration suggested by mean-field theory for an intermediate n value of 100 (Fannon and Nugent, 1996) is nearly $2\mu M$, far higher than the concentrations usually used in experiments (which is of the order of nM). Thus it appears that the high HSPG to CSR ratio observed in cells might be nature's way of tilting the balance in favor of the triad complex, which is more stable and is a likely precursor to effective CSR dimerization and signal transduction.

We have checked the findings from mean-field theory against Monte Carlo simulations of the system using a cellular automaton algorithm. Our results qualitatively support the findings of the mean-field approximation. However, the threshold ligand concentration required for a fair competition between the triad and binary complex is much higher than that predicted by the mean-field rate equations. In fact, for n=1, the ratio of the steady-state concentrations of CSR-FGF-2 and CSR-FGF-2-HSPG is only a little above ½ even at ligand concentrations as high as 5 $\mu M$. We believe that this is caused by a local imbalance in the concentrations of CSR-FGF-2 and HSPG (or FGF-2-HSPG and CSR), which arises because the rates of binding of FGF-2 to CSR and HSPG are significantly different. This imbalance induces the reactions between the bound and unbound species to proceed at a much faster rate than that predicted by mean-field considerations (where such local fluctuations are ignored). A more systematic study of such fluctuations could lead to improved quantitative predictions regarding these effects, but is postponed to future work.

In general, the characteristic time scales in the simulations for reaching the steady state are much shorter than what is found in experiments (example found in Nugent & Edelman, 1992). This could be due to several reasons. Experimental measurements have shown that the diffusion of large proteins (such as CSR and HSPG) on the cell surface often proceeds much slower (by a factor of 10-100) than a simple estimate based on their size would show (Fujiwara *et al.*, 2002). This is usually attributed to steric interactions between large proteins or obstruction to the free diffusion of proteins from the underlying cytoskeleton network. We have not included these effects explicitly in our simple model, but a reduction in the diffusion coefficient by a factor of 1000 (in the absence of pre-coupling) is observed to increase the time scale by a factor of 10, thus bringing it to closer to experimental values (Fig. 9). However, this observation also leaves open the possibility that surface binding processes are governed by a reaction rate which is much slower than the diffusion rate. Also, there is some experimental evidence that all HSPG sites are not equivalent with regard to FGF-2 binding, higher order complex formation, and signaling (Kan *et. al.*, 1999; Knox *et. al.*, 2002). Further, there is evidence that specific HSPGs, syndecan-4 (Simons and Horowitz, 2001) and glycpican-1 (Qiao *et. al.*, 2003), can mediate FGF-2 signaling suggesting that a more complex model where there are both coupling/non-coupling and signaling/non-signaling HSPG may need to be established as more experimental evidence becomes available.

Finally, we add a note on the experimental and biological implications of our work. The typical serum concentration of FGF-2 *in vivo* is low even in pathological conditions (Bairey *et. al.,* 2001; Sezer *et. al.,* 2001). As we have seen from both our mean-field theory and cellular automaton simulations, the high ratio of HSPG to CSR in cells combined with their vastly different affinities for FGF-2 renders pre-coupling of CSR and HSPG essentially irrelevant (even at micro-molar concentrations the triad would be the dominant form of FGF-2-bound CSR). However, our modeling does predict that at super-high ligand concentrations we would see an effect and it would be interesting to check for a possible transition from triad-dominated to binary complex-dominated states for ligand bound CSR.

We acknowledge financial support from the National Science Foundation, Division of Materials Research (grant nos. DMR-0075725 and DMR-0088451), the Jeffress Memorial Trust (grant no. J-594), and the National Institute of Health (grant no. HL56200). MG would also like to acknowledge a fruitful and stimulating discussion with Henk Hilhorst.

**Appendix A**

An estimate of the surface reaction rate constant $\lambda_D$ may be obtained using the well-known Smoluchowski theory (Smoluchowski, von M., 1917). In this theory, two molecules A and B will react with each other if their centers of mass are within a 'reaction radius' a. The effective reaction rate $\lambda_D$ now emerges naturally in terms of the reaction radius a and the 'relative diffusion coefficient' of the species, which is just the sum of the individual diffusion coefficients.

Within Smoluchowski theory, the rate equation for a reaction A+B →C in d spatial dimensions has the general form

$$\frac{d\rho_c(t)}{dt} = K_d(t)\rho_A(t)\rho_B(t) ,\quad\quad\quad (A.1)$$

where $K_d(t)$ is the d-dimensional Smoluchowski factor (effective reaction rate), given by the flux of diffusing particles through an absorbing sphere of radius a centered at the origin, in d spatial dimensions. For our purposes, we need the two-dimensional case (Torney D. C and McConnel, H. M, 1983), which has the form (at sufficiently large times),

$$K_2(t) = \frac{4\pi D}{\log\left(\frac{4Dt}{a^2}\right)} .\quad\quad\quad (A.2)$$

We note that the reaction radius 'a' enters the expression only through a slowly varying logarithmic term. In fact, for our purposes, this logarithmic correction can largely be ignored, which yields the effective reaction rate

$$\lambda_D \approx 4\pi D .\quad\quad\quad (A.3)$$

**Appendix B**

The fraction of CSR existing in the complex state CSR-HSPG is determined as follows. The mean-field rate equation for the reaction kinetics

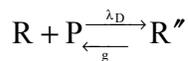

$$R + P \underset{g}{\overset{\lambda_D}{\rightleftarrows}} R''$$

of this complex in the absence of FGF-2 is

$$\frac{dR''}{dt} = \lambda_D RP - gR'',  \tag{B.1}$$

for which the steady state becomes

$$R''_s = \frac{\lambda_D}{g} R_s P_s .  \tag{B.2}$$

The fraction of free CSR and HSPG is related to $R''_s$ through

$$R_s = R_0 - R''_s \quad \text{and} \tag{B.3}$$
$$P_s = P_0 - R''_s .  \tag{B.4}$$

After substituting in eqn. (IIb), and solving the resulting quadratic equation, we find

$$R''_s = \frac{1}{2}\left[ R_0 + P_0 + \frac{g}{\lambda_D} \pm \sqrt{(R_0 + P_0 + \frac{g}{\lambda_D})^2 - 4R_0 P_0} \right] . \tag{B.5}$$

Only one of the roots of this expression gives the correct solution. This is found by applying the limiting condition that $R''_s \to 0$ as $\frac{\lambda_D}{g} \to 0$. Only the negative root satisfies this condition. So it follows that the initial concentration for CSR-HSPG complex prior to ligand absorption is given by

$$r^*(g) = \frac{R''_s}{R_0} = \frac{1}{2R_0}\left[ R_0 + P_0 + \frac{g}{\lambda_D} - \sqrt{(R_0 + P_0 + \frac{g}{\lambda_D})^2 - 4R_0 P_0} \right] . \tag{B.6}$$

Consequently $r^*(g) \to \min(1, n)$ in the opposite limit $\frac{g}{\lambda_D} \to 0$, i.e.,

$R_0 \leq P_0 : R''_s \to R_0, \ R_s \to 0, \ P_s \to P_0 - R_0$ whereas for
$P_0 \leq R_0 : R''_s \to P_0, \ R_s \to R_0 - P_0, \ P_s \to 0.$

**Appendix C**

In this appendix, we study a simpler receptor-ligand problem, when we have only a single type of receptor. In this model, the receptor-ligand complex is stabilized by diffusion-limited reactions with another (unoccupied) receptor molecule. The relevant reaction-diffusion processes are

$$R \xrightarrow{\alpha} R',$$

$$R' \xrightarrow{\beta} R,$$

$$R' + R \xrightarrow{\lambda_D} R''' \quad \text{and}$$

$$R''' \xrightarrow{r} R + R.$$

The corresponding mean-field rate equations can also be written down easily (we again omit the square brackets denoting the concentrations):

$$\frac{dR}{dt} = \beta R' + 2\gamma R''' - \alpha R - \lambda_D R R', \tag{C.1}$$

$$\frac{dR'}{dt} = \alpha R - \beta R' - \lambda_D R R' \quad \text{and} \tag{C.2}$$

$$\frac{dR'''}{dt} = \lambda_D R R' - \gamma R'''. \tag{C.3}$$

These equations are to be supplemented by the normalization relation

$$R + R' + 2R''' = R_0, \tag{C.4}$$

where $R_0$ is the total number of receptor molecules in the cell.

As for the CSR-HSPG system, we define the rescaled dimensionless variables $\tau = \alpha t$, $\beta' = \frac{\beta}{\alpha}$, $\gamma' = \frac{\gamma}{\alpha}$ and $\lambda'_D = \frac{\lambda_D R_0}{\alpha}$. The steady-state relations between the various concentrations are

$$r' = \frac{r}{\beta' + \lambda'_D r}, \tag{C.5}$$

$$r''' = \frac{\lambda'_D r^2}{\gamma'(\beta' + \lambda'_D r)} \quad \text{and} \tag{C.6}$$

$$r = \frac{\lambda'_D - 1 - \beta + \sqrt{(\lambda'_D - 1 - \beta')^2 + 4\beta(\lambda'_D + \eta)}}{2(\lambda'_D + \eta)}; \quad \eta = \frac{2\lambda'_D}{\gamma'}. \tag{C.7}$$

We now search for the different regimes in the parameter space where one type of bound receptor dominates over the other. For example, the condition for $\frac{r'}{r'''} \geq 1$ is

$$\lambda'_D \leq \gamma' \left[ 1 + \frac{3}{\beta' + \gamma'} \right] \quad . \tag{C.8}$$

In terms of the original variables, it becomes

$$\lambda_D \leq \frac{\gamma}{R_0} \left[ 1 + \frac{3\alpha}{\beta + \gamma} \right] \quad . \tag{C.9}$$

This condition says that the triad concentration will decrease as the diffusion coefficient becomes smaller (note that $\lambda_D \approx 4\pi D$ from eqn. (A.3)), which is what one would expect. We have mapped this regime in the steady state diagram shown in Fig. 1B.

# FIGURE CAPTIONS

FIG. 1. Mean-field steady state diagram of the FGF-2-CSR-HSPG system in the absence of direct interaction between CSR and HSPG. The diagram is obtained through numerical solution of the mean-field steady state equations, and shows the regimes where the binary complex CSR- FGF-2 dominates over the triad CSR-FGF-2-HSPG.

FIG. 2. Mean-field steady state diagrams of the CSR-FGF-2-HSPG system and a single receptor system. The lines differentiate between binary-complex dominated (above the line) and triad dominated (below the line) regimes for the two systems. The bold line corresponds to CSR-FGF-2-HSPG system and the thin line corresponds to the single receptor system, which was assumed to have the same absorption and release rates for FGF-2 as CSR in the original model.

FIG. 3. Effect of triad dissociation rate (top line) and the dissociation rate $\beta$ of the binary complex CSR-FGF-2 or HSPG-FGF-2 (bottom line) on the steady state boundary between triad-dominated and binary-complex dominated regimes changes. The effects are similar for the single receptor and two-receptor models.

FIG.4. Effect of pre-formed CSR-HSPG complexes on CSR-FGF-2-HSPG levels. Pre-formed CSR-HSPG complexes were varied via the decay rate g for [FGF-2] of 55nM (·) and 55 $\mu$M (-). Results were obtained by numerical solution of the mean-field steady-state equations.

FIG.5. Effect of pre-formed CSR-HSPG complexes on CSR-FGF-2-HSPG levels as a function of n (n = 1 (bottom), 10 (middle) and 100 (top)) at [FGF-2] = 5.5$\mu$M. Results are obtained by numerical solution of the mean-field steady-state equations.

FIG.6. Formation of CSR-FGF-2-HSPG as a function of time for [FGF-2] = 0.55 $\mu$M (─) and 5.5 $\mu$M (-) with n = 1. Results are obtained by numerical simulation of the cellular automaton model.

FIG.7. Effect of pre-formed CSR-HSPG on steady-state level of CSR-FGF-2-HSPG triads for [FGF-2] = 0.55 $\mu$M (◊) and 5.5$\mu$M (+) for n=1. Results are obtained by numerical simulation of the cellular automaton model. We note that the sensitivity to the presence of pre-formed complexes increases with the ligand concentration. A similar effect is expected also for the single-receptor system discussed in Appendix C (Fig. 2).

FIG.8. Effect of n (ratio of [HSPG] to [CSR]) on the fraction of CSR-FGF-2-HSPG triads for [FGF-2] = 0.55 $\mu$M ($\Diamond$) and 5.5 $\mu$M (+) for r*(g) = 0.05. Results are obtained by numerical simulation of the cellular automaton model. Noticeable effects are seen only at very high ligand concentrations.

FIG.9. Effect of the diffusion coefficient on the level of CSR-FGF-2-HSPG triads as a function of time. Results are obtained by numerical simulation of the cellular automaton model with n=10 and [FGF-2] = 0.55 nM with D = $10^{-11}$ cm$^2$ s$^{-1}$ (—) and $10^{-13}$ cm$^2$ s$^{-1}$ (…). For D=$10^{-9}$ cm$^2$ s$^{-1}$, which is the estimated value in artificial membranes (Kucik *et. al.,* 1991), the time scales are even smaller (data not shown).

**FIGURES**

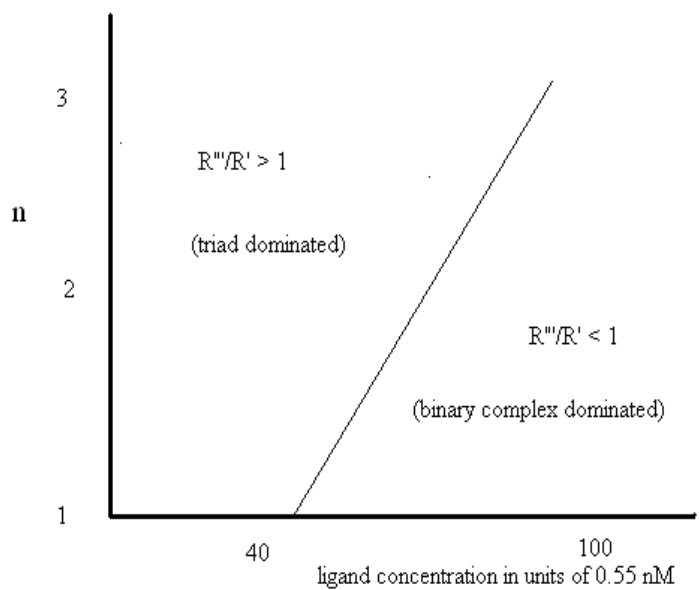

FIG.1

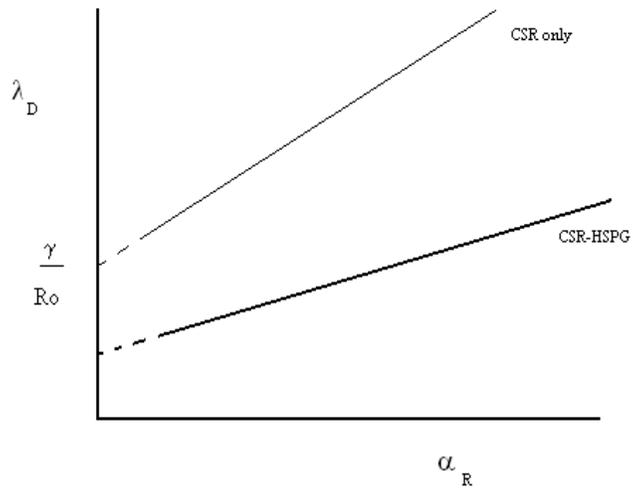

FIG.2

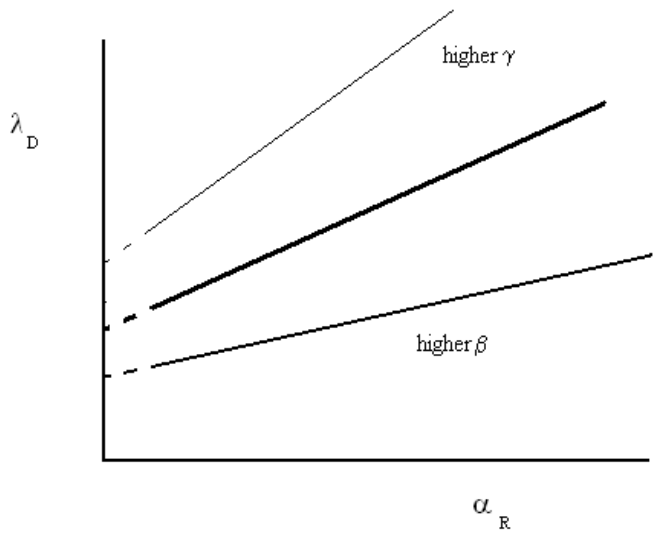

FIG.3

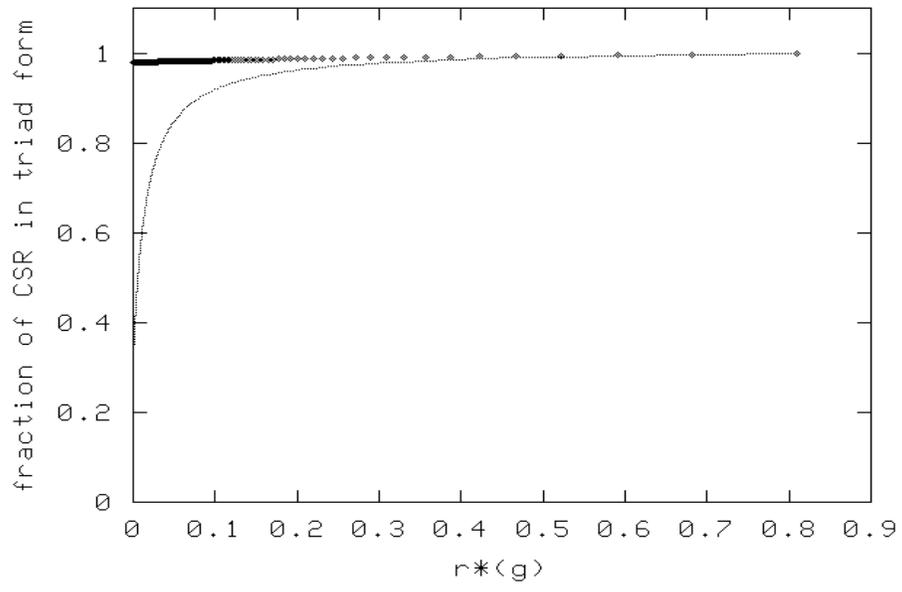

FIG.4

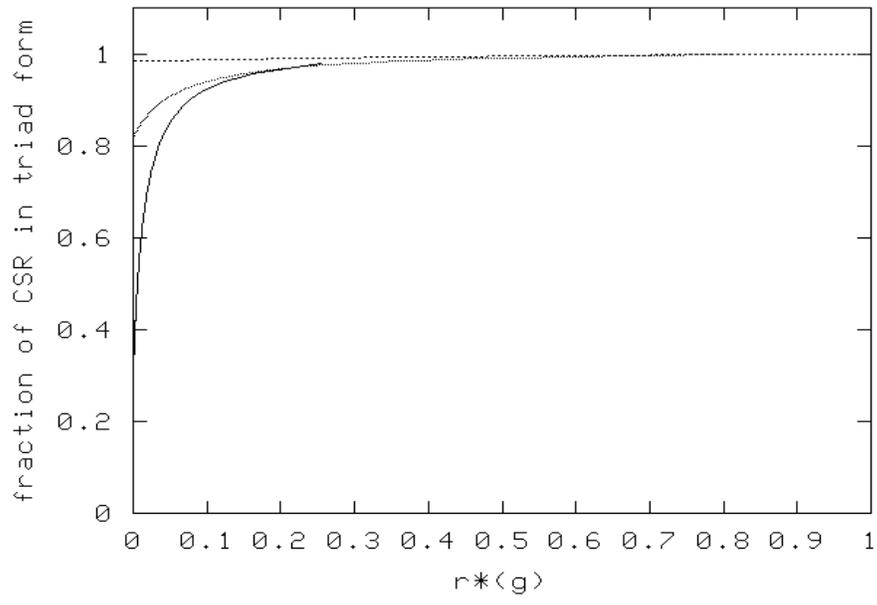

FIG.5

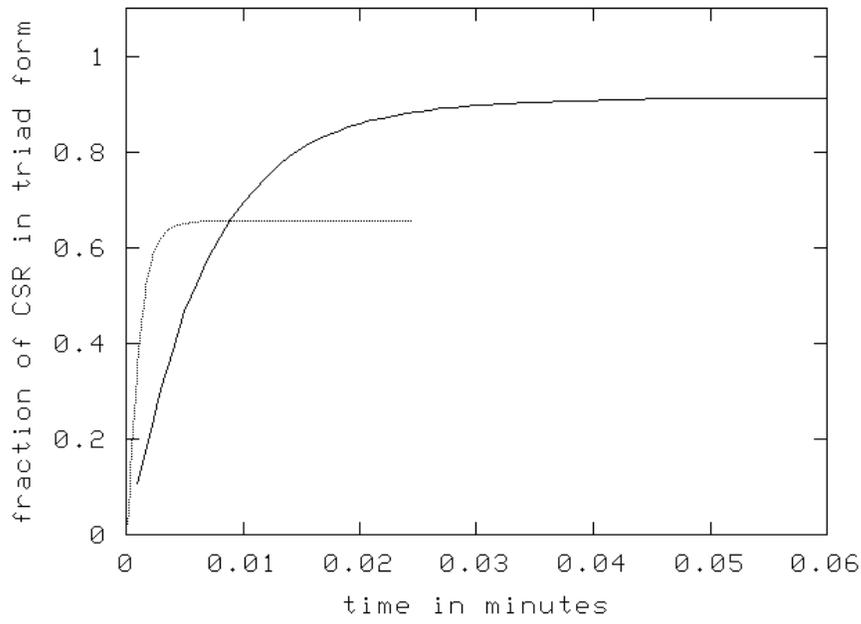

FIG.6

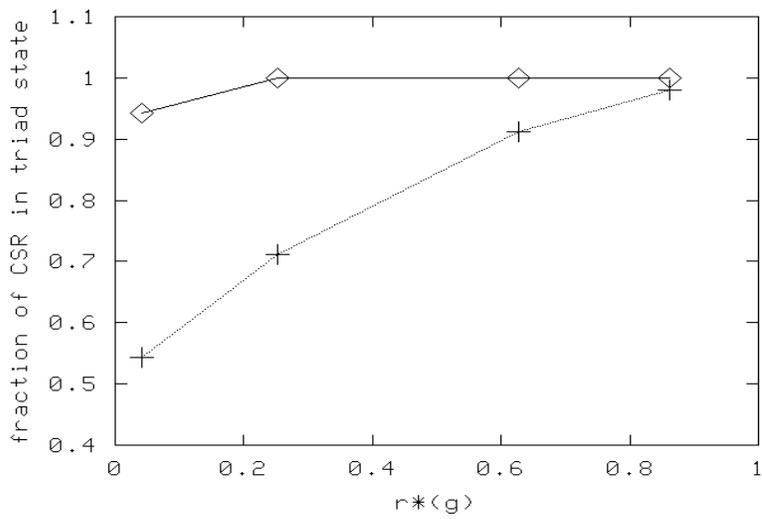

FIG.7

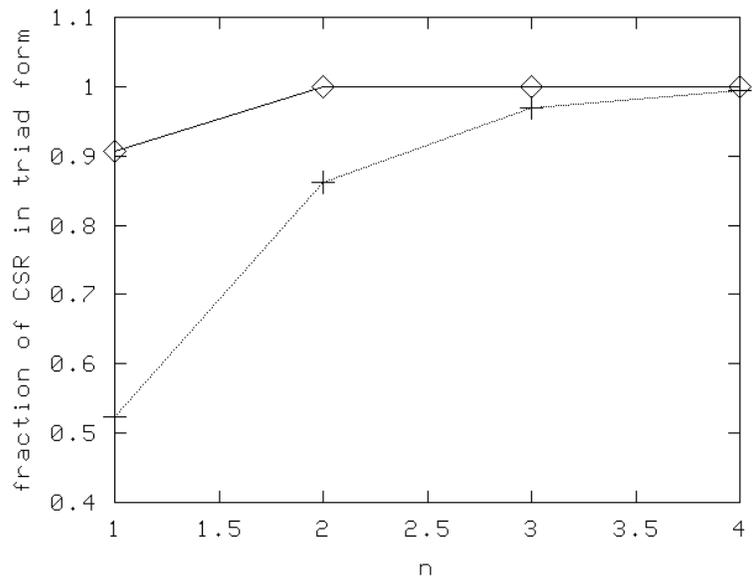

FIG.8

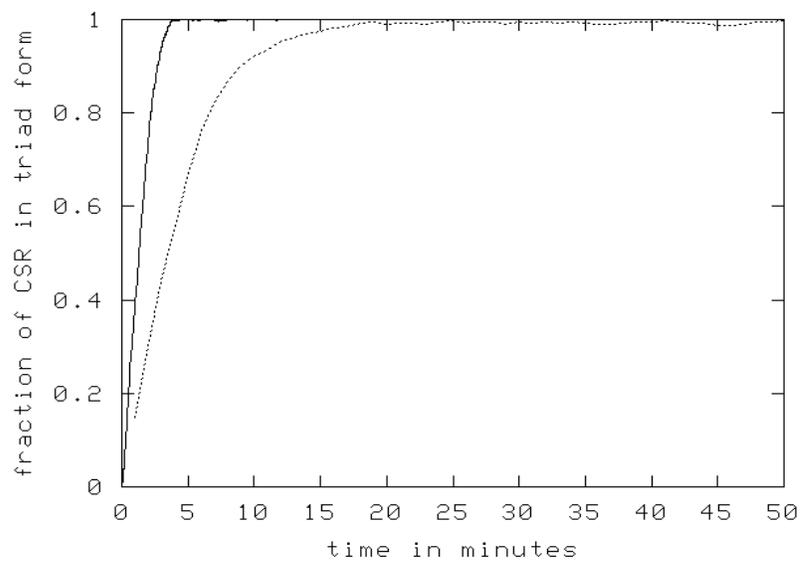

FIG.9.

# TABLES

| $k_{on}^{R}$ | $k_{on}^{P}$ | $\beta_R$ | $\beta_P$ | $\gamma$ |
|---|---|---|---|---|
| $2.27\times10^{8}\,M^{-1}\,min^{-1}$ | $0.9\times10^{8}\,M^{-1}\,min^{-1}$ | $0.048\,min^{-1}$ | $0.095\,min^{-1}$ | $0.003\,min^{-1}$ |

TABLE 1: The experimental values of the various rate constants in our model (Nugent and Edelman, 1992). The binding experiments were performed at a FGF-2 concentration of 0.55nM.